\documentclass[twocolumn,trackchanges]{aastex61}
\usepackage{graphicx}  
\usepackage{enumitem}  

\newcommand{\gc}{$\gamma$\,Cas}
\newcommand{\xr}{X-ray}

\def\gtrsim{\mathrel{\hbox{\rlap{\hbox{\lower4pt\hbox{$\sim$}}}\hbox{$>$}}}}
\def\ltsim{\mathrel{\hbox{\rlap{\hbox{\lower4pt\hbox{$\sim$}}}\hbox{$<$}}}}
\definecolor{gray}{rgb}{0.5,0.5,0.5}

\shorttitle{Ultraviolet activity in $\gamma$\,Cas}
\shortauthors{Smith }

\begin{document}

\title{Ultraviolet activity as indicators of small-scale magnetic fields 
in $\gamma$\,Cassiopeiae}

\author[0000-0003-4240-4794]{Myron A. Smith}
\affil{National Optical Astronomy Observatory 
950 N. Cherry Ave.,
Tucson AZ 85721, USA}

\correspondingauthor{M. Smith}
\email{myronmeister@gmail.com}

\begin{abstract}
The recorded activity of \gc~(B0.5\,IVe) in the ultraviolet
is important to an understanding of the mechanism behind this prototypical Be 
star's high energy activity, especially its hard X-ray emissions.  
Our analysis focuses first on the phasing of ultraviolet and optical light 
curves from different epochs.
The rotational interpretation of the 1.22\,d optical signal is justified
in part on this phasing. Next we detail observations of ``migrating 
subfeatures," which traverse blue to red across line profiles in high quality  
spectra.  These are likely to be proxies for surface magnetic fields, 
which have not been detected yet by spectropolarimetric means in 
rapid rotating Be stars like \gc.~We also analyze the important responses 
of the semi-forbidden resonance S\,IV line at $\lambda$1404.8 to
simultaneous X-ray variations. These results offer indirect support for 
the existence of small-scale magnetic fields on this Be star.

\end{abstract}


\section{Introduction}
\label{intrdn}
  
  $\gamma$\,Cas is the prototypical member of the class of ``classical Be stars"
\citep{2013A&ARv..21...69R} and also the prototype of a subgroup of some 22 
stars that exhibit copious hard \xr\ flux \citep[][]{2018A&A..619A...148N}. 
These fluxes are variable on timescales from a few seconds to a year or longer 
- for a review see \citet[][``SLM16'']{2016AdSpR..58...782S}. Stars of this 
subgroup are evolved on the main sequence and are confined 
to a spectral type range of about O9.7-B1.5. 
We will adopt the following parameters for \gc: an age of at least 20\,Myr, 
a spectral type of B0.5\,IVe, a mass of $\simeq$13-15\,M$_{\odot}$, 
a radius of about 10\,R$_{\odot}$, and a
${\rm v\,sin\,i}$ $\approx$\,441\,km\,s$^{-1}$ \citep{2005A&A..440...305F}.
We view its decretion disk and presumably the star at an 
inclination of 42${\pm 4}^{\circ}$ \citep[][]{2012A&A..545A...59S}.
The last parameters indicate that the star's 
rotation rate is near the critical velocity, a fact also deduced 
from its ${\rm v\,sin\,i}$ being at the upper limit observed for early Be 
stars \citep[][``M15'']{2015ApJ..806...177M}. 
We estimate combined errors for the radius and ${\rm v\,sin\,i}$ as $\pm{20\%}$.
Our inference concerning critical rotation is also consistent with 
the star's periodic photometric signature (with P$_{rot}$ = 
1.21\,d; 0.822 cy\,d$^{-1}$) found from long-term $V$ and $B$ band
monitoring with an Automated Photometric Telescope (APT) system
\citep[][``HS12'']{2012ApJ..760...10H}. This signal had disappeared by 2012
or 2013.

\gc\ is is in a 203.6\,d binary with a near-circular orbit
\citep[][]{2012A&A..537...59N,2012A&A..540...A53S}. 
The companion has a mass of 0.9${\pm 0.4}$ M$_{\odot}$. 
The fact that the secondary's UV flux is $<$0.6\% of the total 
for the system \citep[][]{2017ApJ..843...60W}
demonstrates that any variations in UV spectra, such as discussed 
below, cannot originate from this companion or a disk surrounding it.
Also, although its evolutionary status is otherwise unknown, 
the UV limit does rule out the secondary's being an sdOB star.

 Two general explanations have been put forth for the emission of the 
peculiar hard X-ray emission from this class of variables.  The first is 
that stellar wind or disk wind particles accrete
onto a degenerate secondary, such as a neutron star 
\citep[``NS'';][]{1982ApJ..263...277W,2017MNRAS..465...L119P}
or white dwarf \citep[``WD''][]{1986ApJ..310...L31M,2016ApJ..832...140H}
and convert their liberated potential energy to X-rays.
\citet[][]{2018PASJ..tmp...117T} have contrasted the expected
phenomenalogies for the magnetic and nonmagnetic cases, under the
{\em assumption} of accretion onto some type of WD secondary.

The second scenario, to which we confine ourself herein and for which
the current status of the secondary star is unimportant, 
is the ``magnetic star-disk interaction picture" 
\citep[][``SR99, RS00, RSH'']{1999ApJ..517...866S,2000ApJ..540...474R,
2002ApJ..575...435R}. In this picture small-scale fields emerging from 
the Be star entangle with a toroidal
disk field and reconnect. As they do so, they
accelerate electron beams onto the Be star's surface, where
they produce the ubiquitous, rapid quasi-flares and background
``basal'' flux we observe.  Because internal fossil  magnetic fields in early B 
stars are unlikely to be at work, a key question in the magnetic 
interaction picture is: where do the putative surface fields come from? 

To address this question, M15 pointed to the prediction by Cantiello et al.
\citep[][``CB11a'']{2009A&A..499...279C,2011aA&A..534...A140C,2011bIAUS..272...32C}
that the revised Fe opacities cause a thin equatorial convection zone
to be set up in rapidly rotating massive stars such as \gc.~ They predicted
that nonpermanent circulations are driven in these zones that produce 
so-called $\sigma\omega$ dynamos and local magnetic fields.
Field lines drift to the surface where they emerge.\footnote{We emphasize 
that magnetic fields have not been detected by direct means on 
$\gamma$\,Cas \citep[][]{2018Neiner}. However,these measurements were made
during recent years when the optical bright spots, likely associated with
magnetic fields, had weakened or were absent. Also,
differential rotation can greatly complicate the surface field topology
\citep[e.g.,][]{2003MNRAS..345...1145D} and thus the detection of 
small-scale fields.}
The interaction picture extends this idea by positing the entanglement
of these fields with the disk's, with the ensuing release of X-rays.

This paper is based on an integrated analysis done for all the correlations 
observed in \gc\ between X-ray variations and those in the UV and optical 
domains. 
This approach has led to the extension of facts not comprehensively put 
together to date.
These include:
(1) persistent variations in the UV and optical light curves around the
rotational period and long-cycles observed in the X-ray and optical regimes
(RSH) and (2) the ``migrating subfeatures" in optical and UV line profiles 
(SR99).  We develop these points now so as to facilitate 
planning of optical, \xr~and, in the long-term, UV observations.

\section{Early reports of multi-wavelength correlations in \gc}
\label{pre96}

To recount the history of multi-wavelength correlations discovered 
in \gc,~it is important to include what was observed before recent
generations of UV and X-ray satellites. The first reports were based on 
a simultaneous campaign in 1977 undertaken 
with the {\it Copernicus} satellite 
\citep[][]{1978ApJ..224L...127S,1982PASP..94...157P}.
Let us recall that {\it Copernicus} was able to observe targets 
in the UV and medium X-ray at the same time.

Using the {\it Copernicus} \xr\ camera, \citet{1982PASP..94...157P}
found a  rapid \xr\ ``burst" on 1977 January 27.95; this burst was
confirmed by \citet[][]{1994Polidan} and \citet[][]{1994Parmar}.
At the same time, the {\it Copernicus} UV scanners, V1 and U1, slowly scanned
the  Mg\,II and Si\,IV resonance doublets, respectively. 
One V1 scan disclosed brief emissions in the red wings of both Mg\,II lines. 
Simultaneously, the U1 scanner slowly traversed the Si\,IV doublet.
When it reached $\lambda$1394 it revealed
emission in this line's red wing, just as in the Mg\,II lines.
While these observations were in progress,
Slettebak \& Snow obtained a series of ground-based H$\alpha$ 
spectra and discovered a five-sigma emission ``flare" in this line.

Other flare-like events have been occasionally observed in \gc.~ Similar to 
the 1978 burst of Peters, \citet[][]{1986ApJ..310...L31M} discovered a several 
minute \xr\ ``flare" occurring on 1983 October 31, a circumstance probably
due to a confluence of a few smaller events
\citep[][``SRC'']{1998aApJ..503...877S}.
In the optical,  \citet[][]{1995ApJ..442...812S} discovered 
a ``flare" appearing in the red wing of He\,I $\lambda$6678, just five 
minutes after a spectrum that exhibited no feature. Three minutes after the 
``flare" appeared, another spectrum showed that the feature 
had already weakened but was still visible.
The feature did not reappear in subsequent spectra of the observing night. 
A cosmic ray event as a false signal could be ruled out.

All these events occurred quickly, lasting only minutes.

\section{Recent multi-wavelength correlations} 

Such unexpected events as these were the impetus for a simultaneous campaign 
on \gc\ organized by SRC on 1996 March 14-15 using the {\it Rossi X-ray Timing 
Explorer (RXTE) } and the {\it Goddard High Resolution Spectrograph
(GHRS)}, 
attached to the {\it Hubble Space Telescope.} These 
observations lasted 27 hr and 21.3 hr, respectively. 
The {\it GHRS} spectra were centered on the $\lambda$1394, $\lambda$1403 
complex and covered a total of 36 \AA ngstroms.
Most of the discussion in this paper 
will not focus on the behavior of this doublet.

\subsection{Correspondences in UV and optical time series}
\label{uvopt}

\subsubsection{UV light curves with repeating features}

 To add to the {\it GHRS} spectra, we consulted the 
{\it International Ultraviolet Explorer (IUE)} 
archives to search for high-dispersion, short-wavelength spectral time series
with dense sampling over at least a full rotational period. The archives 
disclose two such records, one in 1982 January and the other in 1996 January.
Their durations were 44 hr and 33 hr, respectively.

Fluxes for wavelengths bins at the blue and red ends of the {\it GHRS} spectra
were coadded to determine a quasi-continuum flux. This step was repeated 
for each spectrum in the time series to determine a ``UVC" light curve. 
We similarly cobinned {\it IUE} {\it Short Wavelength Prime} camera fluxes 
in the echelle orders covering $\lambda\lambda$1200-1260 (except for the
Ly$\alpha$ order) to obtain the corresponding {\it IUE}-based light curves.
The {\it GHRS} curve, originally displayed by 
\citet[][``SRH'']{1998bApJ..508...945S}, is plotted 
as the top curve in Figure\,1.  This curve is rendered with a small 
offset in the ordinate  for visual clarity.
A similar offset was applied to the 1996 January dataset.
Since the SRH paper was published the rotational period of $\gamma$\,Cas 
has been redetermined by the APT optical light curve monitoring 
over 15 years \citep[][``SHV'']{2006aApJ..647...1375S} and HS12. 
These studies resulted in a period P$_{rot}$ = 1.215811 $\pm{0.000030}$\,d.
A recent reanalysis of the APT dataset has shown that the derived
period has a weak dependence on how the prewhitening is performed 
against a stronger 70-80\,d cycle signal.
In fact the derived value would be a few $\sigma$ larger if no prewhitening 
were to be applied to the data \citep[][]{2018Henry}.
In the work described just below we derived a slightly revised value,
P$^{\prime}$$_{rot}$ = 1.21585\,d, from the 
alignment in  phase of optical and UVC features in Fig.\,1. 
This value lies just outside the formal
$\pm{1\sigma}$ limit given by SHV and HS12.\footnote{By extension had we
miscounted cycles by one and accidentally chosen a period corresponding to 
a ${\pm 1}$\,cycle alias, P$^{\prime}$$_{rot}$ 
would be changed by ${\pm 0.0003}$\,d, which is eight times larger
than the period difference given by this revision.} 
Herein we adopt this revised value.

\begin{figure} 
\label{fig1}
\begin{center}
 \vspace*{-.90in}
  \includegraphics*[width=8.5cm,height=11cm,angle=0]{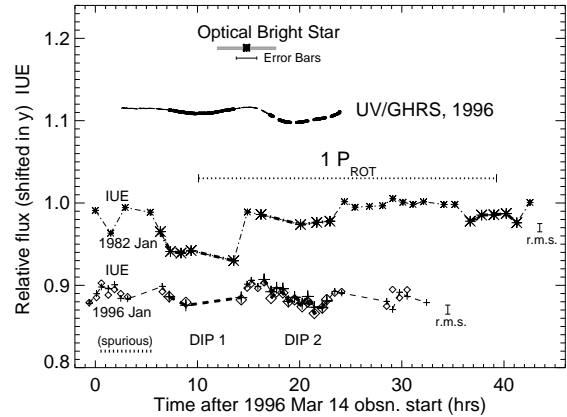}
 \end{center}
 \vspace*{-1.20in}
\caption{The coalignment of the UV dips in the three {\it GHRS} and 
{\it IUE} light curves of \gc,~ 
based on P$^{\prime}$$_{rot}$ = 1.21585\,d. The true dips are shown in bold 
or as large symbols.
The significances of statistically false fluctuations (in range of dotted line)
in the {\it IUE} curves are $\le$2$\sigma$, while the actual dips in
the lower curve are 3.8$\sigma$ and 4.0$\sigma$, respectively. The 
horizontal bar at top 
denotes the phase range of the broad ``bright star" optical maximum of 
the ephemeris of Henry \& Smith (2012) when period-folded with
 P$^{\prime}$$_{rot}$. 
}
\end{figure}

The {\it GHRS} light curve of \gc\ of Fig.\,1 exhibits two principal 
``dip" features of amplitude $\approx$1\% and $\approx$2\%, respectively, 
separated in time by 9.6${\pm 0.5}$\,hr.  SRH noted that two extended {\it IUE} 
light curves of this star showed similar features. 
SRH normalized the UVC echelle fluxes from spectra SWP16127-16193
to create the 1996 January 27-29 UVC curve. We have since improved this
normalization slightly by dividing the far-UV echelle fluxes by sums from
four long-wavelength echelle orders ($m$ = 68-71) instead of just one. 
Although the differences in the relative fluxes for this curve in
Fig.\,1, relative to their original depiction by SRH, are barely 
noticeable, they do serve to better define the noise in flux
from the judged nonvariable stretch of fluxes in the time series.
Note that we dropped the first spectrum of the 1982 series because 
the stellar image was not well centered in the spectrograph entrance aperture.

In Fig.\,1 we have aligned the pairs of dips in the three UVC curves 
that were used to compute the revised P$^{\prime}$$_{rot}$ value.  
The statistical significances of putative dips in the two {\it IUE} curves
shown in the plot were tested by a computer routine, {\it ewcalc.pro,}
the author wrote to assess equivalent width differences from {\it IUE}
spectra using many Monte Carlo realizations, i.e. by first assessing r.m.s. 
noise in the data and then seeing how many simulations are required to
match the total observed flux excursions from the nominal 
level \citep{2006bA&A..459...215S}.
The results of this exercise indicate that a total of three small fluctuations
before time $\approx$6\,hr are not statistically significant. However, the 
two dips in the range ``7-23 hr" appearing  in the 1996 January dataset are
just significant, at 3.8${\pm 0.5}$$\sigma$ and 4.0${\pm 0.5}$$\sigma$.

The 1996 {\it IUE} light curve was not optimal in that the instrument
had to be turned off for what could have been three observations 
during times ``10-13 hr," while the satellite entered the 
{\it South Atlantic Anomaly (SAA)} zone. This duration coincided with the 
predicted time of the first dip, according to the {\it GHRS} record. 
A similar, though less obtrusive, {\it SAA} gap occurred 
during the appearance of the same dip feature in the 1982 light curve. 
Notice that the first and second dips in the 1996 January {\it IUE}
UVC curve may have been similar in strength, in contrast to their relative
sizes in the {\it GHRS} curve 57 days earlier. 
Yet, the first dip was the dominant feature in the 1982 curve, and
one rotation cycle later at $\approx$40\,hr it reappeared much weaker. 
In all, although nominal thermodynamic parameters for 
the absorbing, moderate-sized, ``clouds" responsible for these dips were 
published by SRH, it must be kept in mind that they, or parameters from 
some anisotropic geometry, may not have unambiguous or constant values.

\subsubsection{Bright star optical phase vis \`a vis UVC light curves}

SH12's folded optical light curve for \gc\ (their Fig.\,8) exhibits a broad
``bright-star" plateau. In order to match this waveform at different epochs,
we computed the ephemeris of this feature using the period P$^{\prime}$$_{rot}$ 
and a fiducial $T_{\rm o}$ (computed at Reduced Julian Date = 53000.41\,d) 
for the faint star phase.
Thus, in Fig.\,1 the horizontal bar depicts the phasing
of this plateau, with error bars, during the 1996 March campaign.
In our picture at these times, lasting $\sim$$\frac{1}{3}$ of a rotational
cycle,  an observer will witness two major activity centers on opposite 
limbs of the Be star.  In the moderately precise APT photometry the
centers will appear unresolved as one bright plateau. For the precise UV 
{\it GHRS} dataset they will appear as two distinct, but shallow, minima.

\subsubsection{The flux curve of S\,IV $\lambda$1404.8}

In Figure\,2 we overplot the 1996 March 14-15
{\it RXTE}/X-ray 16\,s fluxes and its derived smoothed curve (dots and dashes,
respectively) and {\it GHRS}/UVC light curve original data (fuzzy dashes).
The latter has been flipped and rescaled to match the extrema of
the smoothed {\it RXTE} curve. As the two curves are almost 
indistinguishable, the correlation between them is excellent. (Note that the
full extent or quality of extended X-ray-UV correlations applies 
at other epochs is not established on the bais of this campaign alone.)

\begin{figure} 
\label{fig2}
\begin{center}
\vspace*{-0.75in}
 \includegraphics*[width=9.0cm,height=13cm,angle=-0]{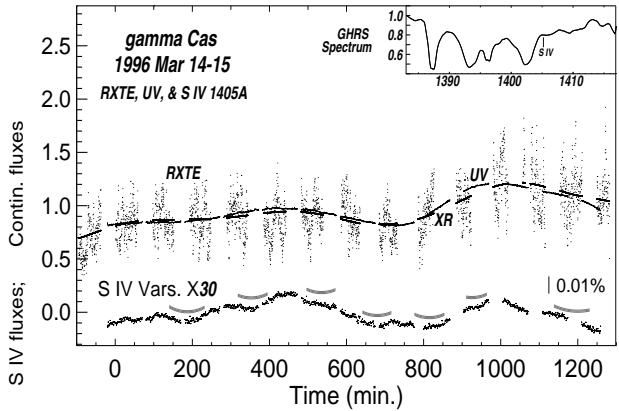}
\end{center}
\vspace*{-1.20in}
\caption{Upper curves in main panel: 
the {\it RXTE} 16\,s (dots) and smoothed data overplotted 
with the inverted UVC continuum light curve. 
The UV data have small r.m.s.  errors and appear here as a fuzzy dashed line. 
The lower curve shows fluxes extracted from the Si\,IV $\lambda$1404.8 line.
The added upward-convex arcs above this curve indicate the times
during which migrating subfeatures are seen in Si\,IV line grayscale.
The top inset shows the position of this S\,IV line in the spectrum. }
\end{figure}

  SLM16 discussed correlations in the flux curves extracted from Fe\,V, 
Si\,IV, and Si\,III lines in the {\it GHRS} spectra in a similar manner 
to the {\it UVC} curve shown in Fig.\,1. They noted that the Fe\,V 
curve was well anti-correlated with the {\it RXTE} curve, even as the
silicon line curves were similarly {\em directly} correlated with it. 
They pointed out that this could best be explained by a shift in 
ionization of the UV-emitting gas 
in response to increased \xr\ emission from a nearby 
\xr\ source(s), thus spatially linking the UV-emitting medium 
to this source. An alternative explanation due to the introduction of 
absorption columns (like those producing the two UVC dips) does not 
serve, since most of these UV lines exhibit {\em decreased} absorption
during these times.

The inset of Fig.\,2 depicts the S\,IV $\lambda$1404.8 line in the 
{\it GHRS} spectrum.  This line is the strongest member of a weak 
resonance multiplet, of which $\lambda$1406.0 is the next strongest. 
Since these transitions are also semi-forbidden, their log\,{\rm gf}
values are some four orders of magnitudes below those of other 
lines in the spectrum (including the Si\,IV doublet). Thus they
probe further into the intervening bodies responsible for the
UVC dips. The S\,IV extraction curve shown exhibits a tight scatter.
Once again the correlation with \xr\ flux is excellent.

\subsection{Spectral migrating subfeatures}

Next we visit an important spectral feature that emerges from the ``grayscale" 
of the {\it GHRS} time series - the grayscale is formed from subtracting 
the mean spectrum from each spectrum of the time series. The result is a 
two dimensional representation of temporal variations. 
The grayscale of our 1996 campaign (see Fig.\,8 of SLM16)
displays prominent gray striations running across 
spectral lines. These have become known as ``migrating subfeatures."

Migrating subfeatures {(\it msf}) are a pattern of blue-to-red absorption
features traversing across the centers of optical and/or UV line profiles 
in two known groups of stars: $\gamma$\,Cas (and perhaps HD\,110432; see 
\citet[][``SB'']{2006ApJ..640...491S}) as one group, and a large group of 
cool, magnetically-active stars of which AB Dor is the prototype.
The {\it msf} phenomenon was first discovered
by \citet[][]{1986ProcASA..6...309R} and ascribed to ``prominences"
forced into corotation by surface magnetic fields. These structures 
are situated typically 1-2 corotation radii above the star
\citep[][``CCRa, CCRb6'']{1989MNRAS..236...57C,1989MNRAS..238...657C}.
Spectropolarimetry has confirmed the presence of multipolar fields on the 
surface of AB\,Dor \citep[][]{2003MNRAS..345...1145D}.\footnote{For purposes 
of comparison to \gc,~\citet[][]{2003MNRAS..345...1145D} found 
a complicated array of surface
fields on AB\,Dor with radial polarities of $\sim$${\pm 500}$\,G.}
The spectra of many stars in this ``AB\,Dor class" suggests that a
complex magnetic topology is a common attribute in this group
\citep[e.g.,][]{2006MNRAS..365...530D,2017AJ..153...152S}.

{\it Msf} were first discovered in optical line profiles of \gc\ by
\citet[][]{1988PASP..100...233Y}  and also observed in ground-based spectra
described by \citet[][]{1995ApJ..442...812S}. 
As implied above, the best sampled observations of them were obtained from 
the {\it GHRS} spectra from the 1996 campaign.
The {\it msf} were found to show an acceleration of +95 km$^{-1}$hr$^{-1}$. 
Also, they were associated mainly with lines of 
less excited ions, e.g., they are typically visible in
S\,IV and Si\,III but not Fe\,V lines. One difference from the {\it msf} of
AB\,Dor spectra is that they are quite short-lived (1 to 1-$\frac{1}{2}$ hr). 
Also, the underlying activity cells are likely to be brighter than the
surrounding stellar surface (CB11a). 
This fact allows them to appear more easily as spectral absorption features.

For the present study we have probed deeper than SR99 and examined the
fainter response of the $\lambda$1406 line as well as the stronger member.
We ran new exoatmospheric {\it CIRCUS} models 
similar to those employed by SR99,\footnote{ The {\it CIRCUS} code, written
by \citep[][]{1996ApJ..470...1144H}, allows one to simulate the spectrum of
a homogeneous foreground gas body of specified size and gas dynamic properties
when seen against the stellar background. The result is 
a narrow absorption at a velocity specified by the assumed corotation
geometry.}
 but we extended the wavelength coverage to 
include the next member, $\lambda$1406.0. From these trials we found 
that an average ratio of their absorption strengths is about 1.5. 
This fixes the optical depth of $\lambda$1404.8 to be 1-3$\times$ the unitary
value that SR99 found. Note this estimate is insensitive to internal 
physical conditions, other than ``microturbulence." Again, be aware that the
$\lambda$1404.8 {\it msf} are absorptions formed in small-scale bodies
 (``cloudlets"). 
The absorption in a cloudlet is determined by the column
lengths and not, as with the larger clouds 
(UV light curve in Fig.\,2), by how close the body is 
to a major X-ray activity cell on the surface. 
To restate the matter, the S\,IV lines respond differently to the two major 
activity centers than they do to the local, smaller centers. 

Clearly, ``cloudlets" 
must be much smaller than the clouds responsible for the UVC dips.
SR99 concluded that it is not likely that a single set of physical parameters 
describes them.  This conclusion emerged from an examination of the reasons for 
the peculiar {\em absence} of {\it msf}  
in the strong Si\,IV doublet (see their Fig.\,10). 
One can simulate this absence by constructing {\it CIRCUS} models 
of the formation of the Si\,IV doublet that include a second, much less 
optically thick, exoatmospheric column that produces {\em deeper line cores} 
than the core computed for the primary, very thick column.  
For the thicker column much more of the line's absorption is
produced in the strongly pressure-broadened wings - and not 
quite so much in line cores. The thin second column obscures the {\it msf} in
the resonance lines that would otherwise appear from the thick column alone.
This counterintuitive result suggests that the absorbing cloudlet is 
heterogeneous in temperature and perhaps in turbulence. After the fact, 
this is not surprising if ``flare" ejecta are in the process of constantly 
being regenerated and cooling on short timescales in a confined prominence-like
structure - our cloudlet.

How high is the elevation of a typical {\it msf} cloudlet? From the case
of AB\,Dor, we know that they can extend to greater elevations 
above the corotation radius, which for the case of \gc\ \,would have to 
be close to the star's surface. We can solve for the elevation of a cloudlet
by determining its distance along the line of sight 
to the star's rotational axis using the
relation given by CCRa:

\begin{equation}
\frac{R_{\mathrm{ax}}}{R_{\mathrm{*}}} \ = \
\frac{c}{\omega \lambda \ vsin\,i \ cos\,\theta} \frac{d\lambda}{dt}.
\end{equation}
Here $\omega$ is the rotational angular frequency, \ d$\lambda$/dt\ the 
{\it msf} acceleration rate of +95 km\,s$^{-1}$hr$^{-1}$ (0.44\,\AA\,hr$^{-1}$),
$\lambda$ is 1404.8\,\AA,\ $\omega$ the star's rotation frequency in 
rad\,hr$^{-1}$, and $\theta$\,=\,45$^{\circ}$ is the assumed stellar longitude.
These parameters, including the revised value for ${\rm v\,sin\,i}$, 
yield a value R$_{elev}$ $\approx$ 0.39\,R$_{*}$.
This estimate would be lower if the activity center associated 
with the cloudlet is situated near the equator.

 Another estimate of the elevation can be made by considering the
projected perpendicular distance traveled by the line of sight connecting the
cloudlet to the stellar surface, starting at the time the foreground cloudlet 
and bright background cell together transit the stellar meridian. 
Once the sightline through the cloudlet moves off the edge of the cell 
(owing to the cloudlet's elevation above the cell), the {\it msf} largely 
disappears. Thus, the elevation is given approximately by the transverse 
velocity of the {\it msf} at meridian passage times the {\it msf}'s 
half lifetime of nearly an hour. 
Assuming a rotational period of 105\,ks, the elevation 
is then $\approx$1.4$\times$10$^{6}$\,km, or 0.2R$_{*}$.
Actually, this estimate is bound to be a lower limit since it does not allow 
for the cell's finite extent.
 
Following SR99, one can estimate the projected area of the 
{\it msf}-forming cloudlets, again using the {\it CIRCUS} program and by 
applying Doppler imaging principles to the problem.
Thus, the absorptions are to be compared with flux of the sector of the 
underlying star with the same radial velocity. 
These computations are mainly sensitive
to the assumed cloudlet temperature (like SR99 we used 18\,000\,K, an
optimum value for the S\,IV lines to form) and assumed a projected
circular shape. Unlike SR99, we recognize now that the {\it msf} features 
are probably narrower than the instrumental resolution (full width 
$\approx$0.144\,Angstroms). We assumed a turbulence of 15\,km\,s$^{-1}$ 
and a line opacity of one. The result is a smaller radius estimate than 
SR99 found,  $\approx$1$\times$10$^{5}$\,km. This is about 30${\pm 10}$ times
the typical size of a pre-flare gas parcel (SRC)  and 
$\sim$$\frac{1}{10}$ the size SRH estimated for the larger ``clouds" 
responsible for the UVC dips.

Because more than one {\it msf} pattern can sometimes be discerned at 
the same time, each in its own phase of its life cycle, several
small centers must be present on the visible hemisphere of the Be star.
However, we emphasize that in our magnetic interaction picture the fields
associated with them must guide electron beams to the surface in order
for \xr\ flux to be observed. The \xr\ flux is visible from this activity
center only during these 
comparatively brief impacts. An active surface cell
 must be \xr\ dormant most of the time because the cooling timescale for
basal flux is at most $\sim$1 hour (SRC).  In our picture for a cell and 
associated cloudlet to be X-ray active over a more extended time requires
successive electron beam bombardments of the cell.  The occurrence of ``flare 
aggregates" in the X-ray light curves suggests that such re-energizations 
happen often.

 Based on extensive {\it IUE} and optical spectroscopic monitoring, 
\citet[][]{2016A&A..594...A56S} and \citet[][]{2018CoSka..48...305S} suggest
that several small magnetic cells are present on the surfaces of two 
semi-evolved O stars ($\lambda$\,Cep, $\xi$\,Per) at a given time.
In their picture several moderate-sized ($\sim$0.2R$_{*}$) magnetic structures,
dubbed ``prominences" appear at the tops of magnetic loops and can be
tracked for several hours as they corotate over the stars' surfaces.
Although this picture is reminiscent of ours, there are contrasts: 
these stars have no decretion disks, nor do they emit copious hard X-rays.

\section{Further discussion} 

\subsection{The wind environment in the ultraviolet} 
\label{cir}
\subsubsection{Ultraviolet dip\MakeLowercase{s}, CIR\MakeLowercase{s}, and 
DAC\MakeLowercase{s}}

High-velocity ``Discrete Absorption Components" (DACs) are ubiquitous 
features in the Si\,IV and C\,IV resonance line profiles in UV 
spectra of early-type Be stars. 
They are particularly prevalent for Be stars observed at intermediate
inclinations \citep[][]{1987ApJ..320...376G}.

 The DACs in the Si\,IV resonance line spectrum of $\gamma$\,Cas 
were first observed by {\it Copernicus} and {\it IUE} 
\citep[e.g.,][]{1982ApJ..253...L39S,1983A&A..268...807H,1994A&A..284...515T}.  
A thorough examination of archival SWP-camera high dispersion spectra of 
\gc~indicates that the features are present to some extent at all times. 
Their strengths can vary over various timescales, even as short as a day.
The {\it GHRS} campaign disclosed that the blue wing profiles could
change appreciably over 1-2 hours.

DACs are believed to be produced from collisions of slow and fast 
wind streams along spiral Corotation Interaction Regions
\citep[CIR;][]{1996ApJ..462...469C}. The multi-streams
are thought to arise from surface ``inhomogeneities" generated by either 
energy leakage from nonradial pulsations (NRP) or small-scale magnetic fields 
over fixed surface regions. The stream-stream collisions trigger shocks that 
propagate back toward their surface origins. Locally in the wind,
they cause a piling up of wind particles and a substantial flattening 
of the  velocity-radius trajectory. As observers we record these as 
absorptions along the far blue wings of the UV resonance lines.

Using the high quality {\it GHRS} spectra from the 1996 March campaign, 
\citet[][]{2000ApJ..537...433C}
found that outward moving DAC elements could be traced back close to
the star's surface and arguably to the two active regions associated with
the UVC dips in Fig.\,1. This result begs confirmation in future UV
observations.

\subsubsection{A possible cause for the UVC dips of \gc}

\citet[][]{2009A&A..499...279C} suggested that magnetic fields created by
subsurface convection cells, lasting as long as several years (CB11a),
can control properties of the wind as 
it emerges as a newly formed hot spot on a rapidly rotating 
massive star.  They predict that the emerging fields would appear as 
hot bright spots that produce comparatively strong wind streams
\citep[][(their Fig. 2)]{2011aA&A..534...A140C,2011bIAUS..272...32C},
which would eventually collide with ambient wind streams. 
We extend this idea to our 
two UVC light curve dips with the following scenario. 
 
\begin{figure} 
\label{gc2014lc}
\begin{center}
\includegraphics*[width= 11cm,angle=00]{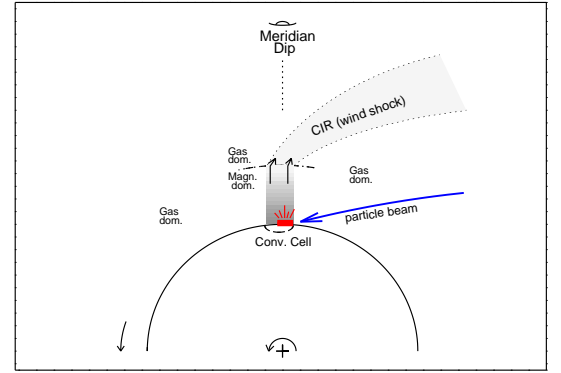}
\end{center}
\vspace*{-0.15in}
\caption{ 
A bird's-eye view of a major magnetic cell
about to be bombarded by an external accelerated electron beam in our scheme.
The slowly accelerating wind (vertical column with variable shading) 
will develop downstream into a CIR.
Because of the star's rotation 
the observer at top sees only briefly a UV dip along the vertical wind column. 
} 
\end{figure}

We posit that the 
emerging magnetic field in flux tubes has a pressure greater than or comparable 
to the local wind pressure and can resist local horizontal motions.
The magnetically controlled wind will then  emerge vertically at first as 
a slowly accelerating, dense column - see Figure\,3.  Overhead, an observer 
will look directly down through a slowly accelerating wind column.
Soon, 
the star will have rotated such that the observer now looks obliquely 
across the column, so that the observed absorption is now much less. 
Thus, in this picture 
the UVC dip is caused by the brief sweeping across our line of sight 
of the wind column. 
SRH modeled the dip absorption as a translucent absorbing cloud and 
found that for reasonable 
parameters its column density must be a few times 10$^{22}$ cm$^{-2}$. 
It remains for field strengths to be discovered to see if in fact they 
can sustain a vertical column of this length before the gas
pressure dominates at higher levels and forces the wind stream to form a spiral.


\subsection{Rotation?}
\label{rotn}

The original motivation for the optical APT monitoring of \gc\ was to 
search for an optical period of just over a day, consistent with the star's 
rotational parameters. Even so, except for the alignment 
of UVC dip features in Fig.\,1, the value 
of the period does not bear directly 
on our arguments for the interaction hypothesis. 

A possible challenge to the rotational interpretation 
of the 1.21\,day period (0.822 cy\,d$^{-1}$) is that
the photometric signal arises instead from NRP. 
This argument is based solely on its near commensurability 
with a 2.48 cy\,d$^{-1}$ frequency observed by the SMEI and BRITE
satellites and the APT \citep[][]{2018Borre,2018Henry}.
The latter frequency likely originates from NRP. 

The strongest argument for a  signal being rotational would be if it 
correlates with an independent variable. In fact this condition is met by the
coincidences of dips at like rotational phase for three UV light curves 
(Fig.\,1). 
Moreover, the 1996 March variations clearly anticorrelates with \xr\ flux
(Fig.\,2).  The phase at which the optical bright star plateau occurs in 
Fig.\,1 would have to be seen as yet another coincidence. 

SRH's  spectrophotometric analysis of the 1996 January {\it IUE} data
demonstrated that the dips are caused by foreground cool structures of 
temperature $\ltsim$10,000\,K - a conclusion confirmed by the variations 
of the spectral line strengths, such as of the S\,IV line (Fig.\,2).  
Indeed, the strong wavelength dependence of the dip amplitudes in the UV 
contrasts sharply with the wavelength-independent character at optical 
wavelengths.  The UV wavelength dependence is consistent with the foreground 
absorption by a cool cloud but not with NRP.

An additional argument for the optical signal being rotational
comes from the contrast in its {\em waveform} from those of NRP modes.
SRC noted that close scrutiny of the {\it GHRS} light curve shows that the 
'`primary dip" at 20 hr in Fig.\,1 
is in fact the superposition of two semi-resolved dips, which introduces 
an extra wiggle in this light curve.  This is not consistent with the
NRP variations, which exhibit symmetrical if not nearly sinusoidal waveforms. 
For example, the waveform for the 2.48\,cy\,d$^{-1}$ variation clearly 
appears sinusoidal \citep[][]{2018Borre,2018Henry}. 
In contrast, not only is the waveform of the optical light curve of 
$\gamma$\,Cas markedly asymmetrical, but the {\em sense} of its asymmetry 
reversed itself in late 2003 (compare Fig.\,7 of SHV with Fig.\,8 
of HS12). It is difficult to understand how this change could arise 
from the action of a stable  NRP mode or to interferences with another mode(s)
- or for that matter how a waveform asymmetry can be sustained over
a several year-long time interval. Rather, it seems that the distribution
of activity centers on the star's surface has changed.

 Contrariwise, the question arises: 
could the 2.48\,cy\,d$^{-1}$ signal be due to the presence
of a third major activity cell, separated by 120$^{\circ}$ in longitude from 
the other two?  For this to be the case, one would expect a 
clear, separated third dip (or a portion of one) in each of the three UV
light curves of Fig.\,1, and yet there are none in any of them. 
Moreover, if there were to be
an NRP mode in a 3:1 resonance with rotation, its excitation would require 
a perturbing agent that is nearly stationary in the star's inertial frame. 
The rather distant secondary star might be such an agent, but to assign 
it as such is at this stage would be dubious and {\it ad hoc.}

From these arguments, we take the rotational interpretation for the 
0.822 cy\,d$^{-1}$ signal to be well-grounded.
This signal is likely to have a physical cause that is independent
of the 2.48\,cy\,d$^{-1}$ oscillation.

\section{Desiderata for future work}

From the foregoing the case for near-term optical and 
future UV monitorings can be advanced as follows:

\begin{enumerate}

\item Future optical monitoring of the light curve of \gc\ should search
for a {\em reoccurrence} of a near-1.21\,d period. Should it reappear, one 
cannot expect a phase match with respect to the old ephemeris since an active 
magnetic cell that causes it could appear on the Be star surface at another 
longitude.  It would be illuminating to see if the hypothetical reborn period 
has been slightly altered, consistent with a spot appearing at a new latitude 
and signaling differential rotation.


\item A future UV space telescope should be employed to 
understand the nature of the UV light curve dips. This includes the 
question of whether their reoccurrences will reappear at the period 
P$^{\prime}$$_{rot}$.  
A related issue is whether two or
even more active centers are robustly present on the Be star's surface. 
Observations of brief soft-\xr\ ``dips" \citep[][]{2016ApJ..832...140H}
may also serve this purpose.

\item  Even a small UV spectroscopic telescope can record 
spectra of \gc\ at high enough resolution to study the variability of 
DACs in the Si\,IV doublet. An important question to resolve 
is whether their strengths correlate with the UVC dip amplitudes.

\end{enumerate}

\vspace{-0.30in}

\acknowledgments

The author is pleased to thank Greg Henry for his analyses leading to
the rotational period and Drs. Raimundo Lopes de Oliveira and Chris Shrader
for suggestions that have improved this paper. 
Many comments by the referee have materially added to its content. 
We are very grateful to Dr.  Christian Motch for organizing the 2018
Strasbourg conference ``The \gc~phenomenon in Be stars," which led to this work.

\end{document}